\documentclass[12pt]{article}

\def\lsim{\:\raisebox{-0.5ex}{$\stackrel{\textstyle<}{\sim}$}\:}
\def\gsim{\:\raisebox{-0.5ex}{$\stackrel{\textstyle>}{\sim}$}\:}

\begin{document}

\renewcommand{\thefootnote}{\fnsymbol{footnote}}

\mbox{ } \\[-1cm]
\mbox{ }\hfill TUM--HEP--452/02\\%[-1mm] 
\mbox{ }\hfill hep--ph/0202072\\%[-1mm] 
\mbox{ }\hfill \today\\%[-1mm]

\begin{center}
  {\Large\bf Production of ultra--energetic cosmic rays through the decay of
    super--heavy X particles} \\[8mm]
           
Cyrille Barbot and Manuel Drees \\[4mm]

{\it Physik Dept., TU M\"unchen, James Franck Str., D--85748
 Garching, Germany} \\[1mm]
\end{center}

\bigskip
\bigskip 
\bigskip 

\begin{abstract}

\vskip 0.5cm

\noindent
%PACS number(s): 11.30.Pb, 11.30.Er

We present a new and complete numerical analysis of the decay of
super--heavy $X$ particles, assumed to be the origin of cosmic rays
with energy beyond the GZK cut--off. The decay of $X$ initiates a
``parton shower'', where we include all degrees of freedom contained
in the Minimal Supersymmetric Standard Model (MSSM). Since at energies
near $M_X$ all gauge couplings are of similar magnitude, we include
all of them, as well as third generation Yukawa couplings. Technically
the shower development is described through the DGLAP evolution of the
relevant fragmentation functions (FFs). We also carefully treat the
decay of the superparticles as well as heavy SM particles created in
the shower. Nonperturbative physics is parameterized through the input
values of the FFs, which we take from the literature. The final result
is the the complete spectrum of all stable particles at the very end
of the shower: protons, electrons, neutrinos, photons and neutralinos, for
an energy range from $10^{-7} M_X$ to $M_X$. In particular, the flux
of high--energy neutralinos is sizable; it might serve as ``smoking
gun'' signature for this kind of scenario.

\end{abstract}

\newpage

%%%%%%%%%%%%%%%%%%%%%%%%%%%%%%
\section{Introduction}
\label{sec:introduction}
%%%%%%%%%%%%%%%%%%%%%%%%%%%%%%
\setcounter{footnote}{1}

The origin of the Ultra Energetic Cosmic Rays (UHECR) is still an
enigma for scientists. We neither know how and where they are
produced, nor what they are. Photons with energy $E_\gamma \geq
10^{15}$ eV can be absorbed on the cosmic microwave background (CMB)
through $e^+e^-$ pair production; at $E_\gamma \sim 10^{20}$ eV, the
main energy loss comes through interactions with the radio background.
Electrons in addition loose energy through synchrotron radiation in
intergalactic and galactic magnetic fields, the strengths of which are
not known very well. Protons and heavy nuclei of energies
above a few times $10^{19}$ eV loose energy through interactions with
the CMB; one would thus expect their spectrum to cut off at that
energy (the so called GZK cut-off \cite{GZK:1},\cite{GZK:2}).
However, a few events have been observed with energies beyond this
cut--off \cite{AGASA}. This means that the UHECR should either have
been created within a few interaction lengths from the Earth, i.e. at
a distance $\lsim 100$ Mpc, or they should have been produced at even
higher energies than are observed now. The first possibility is in
principle consistent with the ``classical'' or ``bottom-up''
explanation for the origin of UHECR, based on ideas already developed
to explain the spectrum of CR at lower energies, i.e. through
acceleration by electromagnetic fields; however, no object in our
cosmological neighborhood is known which has a sufficiently strong
magnetic field extending over a sufficiently large volume. The second
possibility would require the existence of very massive and very
long--lived $X$ particles, with mass $m_X \gg 10^{20}$ eV and lifetime
close to or larger than the age of the Universe. In these ``top-down''
scenarios the observed UHECR are the stable decay products of these
ultra--massive particles (for reviews, see
\cite{reviewSigl},\cite{reviewSarkar}).

In this article we consider a generic top--down scenario.  We provide
an improved model of the decay of an $X$ particle, which is
independent of the origin and the nature of this particle. In leading
order in perturbation theory $X$ decays into a small number of very
energetic particles. However, in reality this should be considered
to be the starting point of a ``parton shower'', analogous to the
formation of hadronic jets in decays of $Z$ bosons studied at LEP. The
existence of an energy scale $m_X$ much above the weak scale requires
supersymmetry (SUSY) to stabilize the electroweak scale against
radiative corrections. We therefore study this shower in the framework
of the MSSM, which is the simplest potentially realistic
supersymmetric extension of the Standard Model (SM). In contrast to
previous works
\cite{Berezinsky:2000},\cite{Coriano:2001},\cite{Sarkar:2001}, we
considered all gauge interactions as well as third generation Yukawa
interactions, rather than only SUSY--QCD; note that at energies above
$10^{20}$ eV all gauge interactions are of comparable strength. The
inclusion of electroweak gauge interactions in the shower gives rise
to a significant flux of very energetic photons and leptons, which had
not been identified in earlier studies. Moreover, we carefully
modeled decays of all unstable particles. As a result, our
treatment is the first that respects energy conservation.

In the next Section we describe our treatment of the parton shower in
slightly more detail. In Sec.~3 we show some numerical results for the
expected spectra of stable particles, and compare it with previous
works. Finally, a brief summary and some conclusions are presented in
Sec.~4.

%%%%%%%%%%%%%%%%%%%%%%%%%%%%%%%%%%%%%%%%%%%%%%
\section{Calculation of the fragmentation functions}
\label{sec:cascade}
%%%%%%%%%%%%%%%%%%%%%%%%%%%%%%%%%%%%%%%%%%%%%%

Consider the two--body decay of an ultra--massive $X$ particle of mass
$m_X \gg 10^{20}$ eV into a $q\bar{q}$ pair, in the framework of the
MSSM. This triggers a parton shower, which can be understood as
follows. The $q$ and $\bar q$ are created with initial virtuality $Q_X
\sim \frac{M_X}{2}$. Note that $Q_X^2 > 0$. The initial particles can
thus reduce their virtuality, i.e. move closer to being on--shell
(real particles), by radiating additional particles, which have
initial virtualities $< Q_X$.These secondaries then in turn initiate
their own showers. The average virtuality and energy of particles in
the shower decreases with time, while their number increases (keeping
the total energy fixed, of course). As long as the virtuality $Q$ is
larger than the electroweak or SUSY mass scale $M_{\rm SUSY}$, all
MSSM particles can be considered to be massless, i.e. they are all
active in the shower. However, once the virtuality reaches the weak
energy scale, heavy particles can no longer be produced in the shower;
the ones that have already been produced will decay into SM particles
plus the lightest superparticle (LSP), which we assume to be a stable
neutralino. Moreover, unlike at high virtualities, at scales below
$M_{\rm SUSY}$ the electroweak interactions are much weaker than the
strong ones; hence we switch to a pure QCD parton shower at this
scale. At virtuality around 1 GeV, nonperturbative processes cut off
the shower evolution, and all partons hadronize. Most of the resulting
hadrons, as well as the heavy $\tau$ and $\mu$ leptons, will
eventually decay. The end product of $X$ decay is thus a very large
number of stable particles: protons, electrons, photons, the three
types of neutrinos and LSPs; we define the FF into a given particle to
include the FF into the antiparticle as well, i.e. we do not
distinguish between protons and antiprotons etc.

Note that at most one out of these many particles will be observed on
Earth in a given experiment. This means that we cannot possibly
measure any correlations between different particles in the shower;
the only measurable quantities are the energy spectra of the final
stable particles, $d \Gamma_X / d E_P$, where $P$ labels the stable
particle we are interested in.\footnote{Of course, this just describes
the spectrum of particle $P$ at the place where $X$ decays. The
spectrum can be changed dramatically by interactions with the
interstellar and intergalactic medium \cite{reviewSigl}. We will not
address this issue in this paper.} This is a well--known problem in
QCD, where parton showers were first studied. The resulting spectrum
can be written in the form \cite{QCDrev}
\begin{equation} \label{def_ff}
\frac {d \Gamma_X} {d x_P} = \sum_I \frac {d \Gamma(X \rightarrow I)}
{d x_I} \otimes D^P_I(\frac{x_P}{x_I}, Q_X^2),
\end{equation}
where $I$ labels the MSSM particles into which $X$ can decay, and we
have introduced the scaled energy variable $x = 2 E / m_X$; for a
two--body decay, $d \Gamma(X \rightarrow I) / d x_I \propto
\delta(1-x_I)$. The convolution is defined by $f(z) \otimes g(x/z) =
\int_{x}^{1} f(z)g\left(\frac{x}{z}\right)\,\frac{dz}{z}$.

All the nontrivial physics is now contained in the fragmentation
functions (FFs) $D^P_I(z,Q^2)$. Roughly speaking, they encode the
probability for a particle $P$ to originate from the shower initiated
by another particle $I$, where the latter has been produced with
initial virtuality $Q$. This implies the ``boundary condition''
\begin{equation} \label{boundary}
D^J_I(z,m_J^2) = \delta_I^J \cdot \delta(1-z),
\end{equation}
which simply says that an on--shell particle cannot participate in the
shower any more. For reasons that will become clear shortly, at this
stage we have to include all MSSM particles $J$ in the list of
``fragmentation products''. The evolution of the FF with increasing
virtuality is described by the well--known DGLAP equations
\cite{QCDrev}:
\begin{equation} \label{dglap}
\label{e1}
\frac{dD_I^J} {d\ln(Q^2)} (x,Q) =
\sum_K \frac{\alpha_{KI} (Q^2)} {2\pi} P_{KI}(z) \otimes
D_K^J(x/z,Q^2),
\end{equation}
where $\alpha_{KI}$ is the coupling between particles $I$ and $K$, and
the splitting functions $P_{KI}$ describes the probability for
particle $K$ to have been radiated from particle $I$. As noted
earlier, for $Q > M_{\rm SUSY} \sim 1$ TeV we allow all MSSM particles
to participate in the shower. Since we ignore first and second
generation Yukawa couplings, we treat the first and second generations
symmetrically. $I,J,K$ in eq.(\ref{dglap}) thus run over 30 particles:
6 quarks $q_L, u_R, d_R, t_L, t_R, b_R$, 4 leptons $l_L, e_R, \tau_L,
\tau_R$, 3 gauge bosons $B, W, g$, the two Higgs fields of the MSSM,
and all their superpartners. Here we sum over all color and $SU(2)$
indices (i.e., we assume unbroken $SU(2)$ symmetry), and we ignore
violation of the CP symmetry, so that we can treat the antiparticles
exactly as the particles. All splitting functions can be derived from
those listed for SUSY--QCD in \cite{Jones}; explicit expressions will
be given in a later paper. The starting point of this part of the
calculation is eq.(\ref{boundary}) at $Q = M_{\rm SUSY}$. This leads
to $30 \times 30$ FFs $D_I^J$, which describe the shower evolution
from $Q_X$ to $M_{\rm SUSY}$.

At scales $Q < M_{\rm SUSY}$ all interactions except those from QCD
can be ignored. Indeed, at scales $Q < Q_0 \simeq$ 1 GeV QCD
interactions become too strong to be treated perturbatively. leading
to confinement of partons into hadrons. This nonperturbative physics
cannot be computed yet from first principles; it is simply
parameterized, by imposing boundary conditions on the
$D_i^h(z,Q_0^2)$, where $h$ stands for a long--lived hadron and $i$
for a light parton (quark or gluon). Here we used the results of
\cite{Poetter}, where the FFs of partons into protons, neutrons, pions
and kaons are parameterized in the form $Nx^{\alpha}(1-x)^{\beta}$,
using fits to LEP results. Note that these functions are only valid
down to $x = 0.1$; for smaller $x$, mass effects become relevant at
LEP energies. However, at our energy scale these effects are still
completely irrelevant, even at $x=10^{-7}$. We chose a rather simple
extrapolation at small $x$ of the functions given in \cite{Poetter},
of the form $N'x^{\alpha'}$. We computed the coefficients $N'$ and
$\alpha'$ by imposing energy conservation. We had to use the set of
NLO FF given in \cite{Poetter}, although our evolution equations are
only written at the leading order; the LO set violates energy
conservation badly, especially for the gluon FF. This choice renders
difficult the comparison with previous results \cite{Sarkar:2001}
\cite{Berezinsky:2000}, where the LO set was used. Apart from this
difference, our results for the pure QCD case or for the SUSY QCD
evolution seem to agree rather well. Starting from these modified
input distributions, and evolving up to $Q = M_{\rm SUSY}$ using the
pure QCD version of eq.(\ref{dglap}), leads to FFs $D_i^h(x,M^2_{\rm
SUSY})$ which describe the QCD evolution (both perturbative and
non--perturbative) at $Q < M_{\rm SUSY}$.

Finally, the two calculations have to be matched together. First we
note that ``switching on'' $SU(2)$ and SUSY breaking implies that we
have to switch from weak interaction eigenstates to mass eigenstates.
This is described by unitary transformations of the form $D^S_I =
\sum_J |c_{SJ}|^2 D^J_I$, with $\sum_S |c_{SJ}|^2 = \sum_J |c_{SJ}|^2
= 1$; here $S$ stands for a physical particle. We used the ISASUSY
code \cite{Isasusy} to compute the SUSY mass spectrum and the
$|c_{SJ}|$ corresponding to a given set of SUSY parameters. The decay
of all massive particles $S$ into light particles $i$ is then
described by adding $\sum_S D^S_I \otimes P_{iS}$ to the FFs $D^i_I$
of light particles $i$. We assumed that the $x-$dependence of the
functions $P_{iS}$ originates entirely from phase space. In this
fashion each massive particle $S$ is distributed over massless
particles $i$, with weight given by the appropriate $S \rightarrow i$
decay branching ratio. Note that this step often needs to be iterated,
since heavy superparticles often do not decay directly into the LSP, so
that the LSP is only produced in the second, third or even fourth
step. All information required to model these cascade decays have
again been taken from ISASUSY. The effects of the pure QCD shower
evolution can now be included by one more convolution, $D^h_I(Q_X) =
\sum_i D^h_i(M_{\rm SUSY}) \otimes D^i_I(Q_X)$. Finally, the decay of
long--lived but unstable particles $\mu, \tau, n, \pi, K$ has to be
treated; this is done in complete analogy with the decays of particles
with mass near $M_{\rm SUSY}$.

Note that the linearity of the evolution equation (\ref{dglap})
allowed us to factorize the problem in the fashion described above. We
integrate these equations numerically using the Runge--Kutta
method. The various FFs are modeled as cubic splines, with about 100
$x$ values distributed logaritmically between $x = 10^{-7}$ and $0.5$,
and again logarithmically in $1-x$ for $x$ between 0.5 and 1.  This
allows us to compute the CR spectrum down to energies of the order of
$10^{18}$ eV even if $M_X = M_{\rm GUT} \simeq 10^{16}$ GeV.  We
should stress that a good test of our algorithm was the possibility of
checking energy conservation at all steps. We found that the ``loss''
of energy due to numerical approximations (the loss to particles with
$x < 10^{-7}$ being negligible) never exceeds a few percent.

%%%%%%%%%%%%%%%%%%%%%%%%%%%%%%%%%%%%%%%%%%%%%%%%%%%%%%%%%
\section{Results and discussion}
\label{sec:results}
%%%%%%%%%%%%%%%%%%%%%%%%%%%%%%%%%%%%%%%%%%%%%%%%%%%%%%%%%
\setcounter{footnote}{0}

We are now ready to present results for the FFs of any particle $I$ of
the unbroken MSSM produced at high virtuality through the decay of $X$
into one of the 7 stable particles $P$. This requires $30\times 7 =
210$ FFs for any set of SUSY parameters, which enable us to predict
any CR spectrum near the source for any decay mode of $X$ into MSSM
particles. We saw earlier that for 2--body decays of $X$, the spectrum
of $P$ is directly given by the relevant FFs; eq.(\ref{def_ff}) shows
that any $N-$body decay \cite{Sarkar:2001} can be treated with just
one more convolution. We computed all 210 FFs for several sets of soft
SUSY breaking parameters, but we found that most features depend very
little on this choice. Here we only give results for the ``low
$\tan\beta$ gaugino region'' (ratio of Higgs vevs $\tan \beta = 3.6$,
gluino mass $m_{\tilde{g}} \sim 400$ GeV, supersymmetric Higgs mass
parameter $\mu \sim 500$ GeV, slepton masses around 140 GeV, squark
masses around 300 GeV, CP--odd Higgs boson mass $m_A = 180$ GeV and
trilinear soft breaking parameter $A_t \sim 1$ TeV). As usual, we show
results for $x^3\times D_I^P (x,M_X)$, appropriate for comparing with
the flattening of the observed spectrum (beyond the region of the knee
characterized by a power law spectrum with power $n \simeq -2.7$). We
take $M_X = 10^{16}$ GeV, as appropriate for a GUT interpretation of
the $X$ particle. According to ref.\cite{Fodor}, such a large value of
$M_X$ is compatible with existing data if most UHECR originate at
cosmological distances. Since the FFs evolve only logarithmically with
$Q$, the final results for $M_X = 10^{12} - 10^{13}$ GeV \cite{Birkel,
Sarkar:2001} would not differ too much from the ones shown here, if
they are expressed in terms of the scaling variable $x$.

\begin{figure}[h]
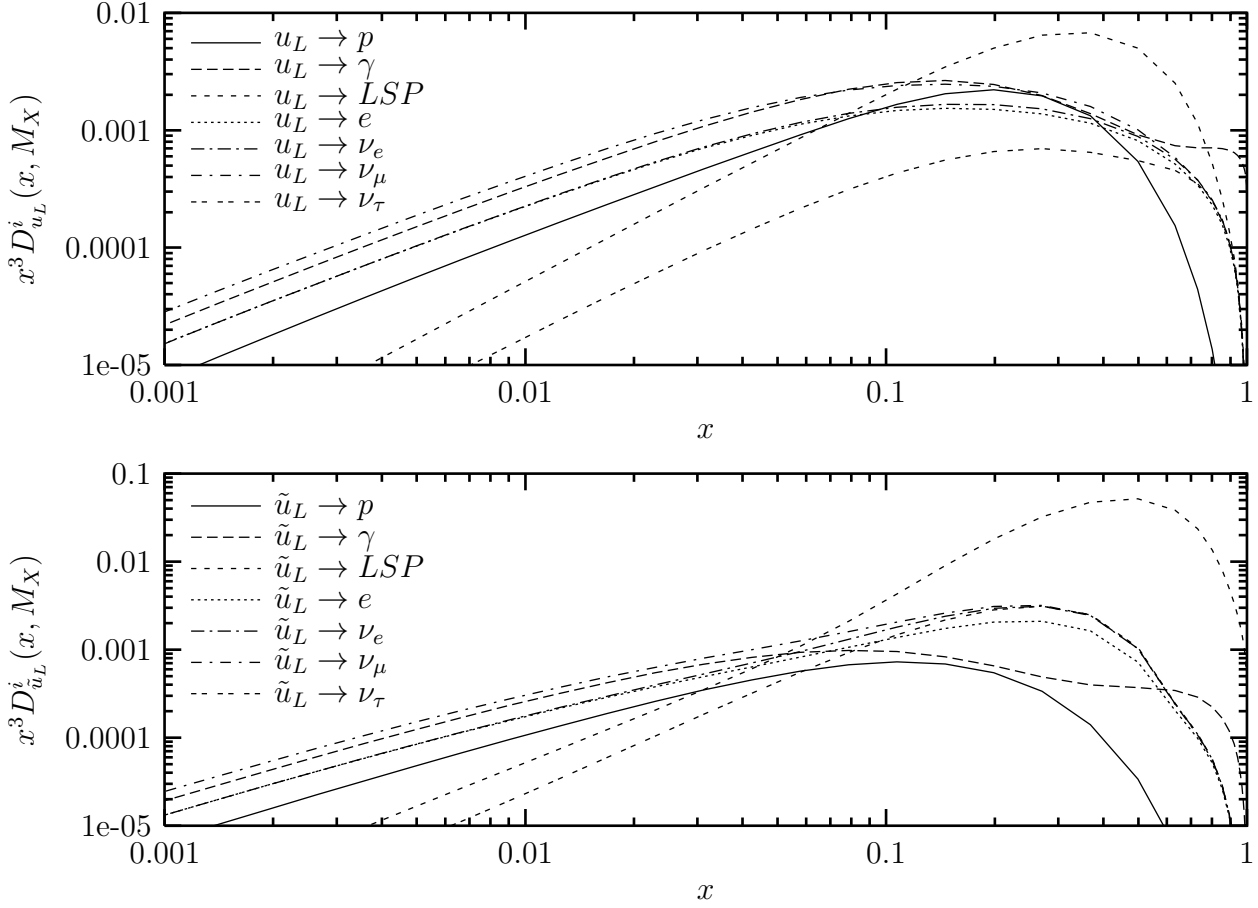
 
\setlength{\unitlength}{1cm}
\begin{minipage}[h]{6.5cm}
\input{Low_G_uL.tex}
\end{minipage}

\noindent
\begin{minipage}[h]{6.5cm}
\input{Low_G_uL_.tex}
\end{minipage}\hfill%
\caption{Fragmentation functions of a first generation $SU(2)$ doublet
quark (top) and a squark (bottom) into stable particles. }
\label{quarks}
\end{figure}

Results for the fragmentation functions of first generation $SU(2)$
doublet (s)quarks are shown in fig.~\ref{quarks}.\footnote{Recall that
  we assume exact $SU(2)$ invariance for $Q > M_{\rm SUSY}$ and treat
  first and second generation (s)quarks symmetrically, so that $q_L$
  is actually composed of $u_L, d_L, c_L$, $s_L$ and their
  antiparticles, and similarly for $\tilde q_L$.} For small $x \lsim
0.01$, the FFs turn out to be very similar for these two cases.
However, at large $x \gsim 0.1$ some differences appear. In particular,
an initial squark produces many more hard LSPs than a quark does; in
the former case, LSPs carry $\sim 25$\% of the original squark energy,
while they only carry $\sim 7$\% of the energy of an initial quark.
Similarly, an initial squark produces a significantly larger flux of
very energetic neutrinos, since neutrinos are frequently produced in
sparticle decays; for $\tan\beta \gg 1$ the majority of these very
energetic neutrinos would be $\nu_\tau$, but for the given small value
of $\tan\beta$ all three neutrino species are produced with equal
abundance. In contrast, in case of an initial quark the three neutrino
fluxes are of similar size only at very large $x$; these neutrinos
come from the radiation of $SU(2) \times U(1)_Y$ gauge bosons early in
the shower, and their subsequent decay. At smaller values of $x$ the
$\nu_\tau$ component drops quickly relative to the $\nu_\mu$ and
$\nu_e$ components, since only the latter are produced in the decays
of light mesons, in the ratio of roughly 2:1.

\begin{figure}[h]
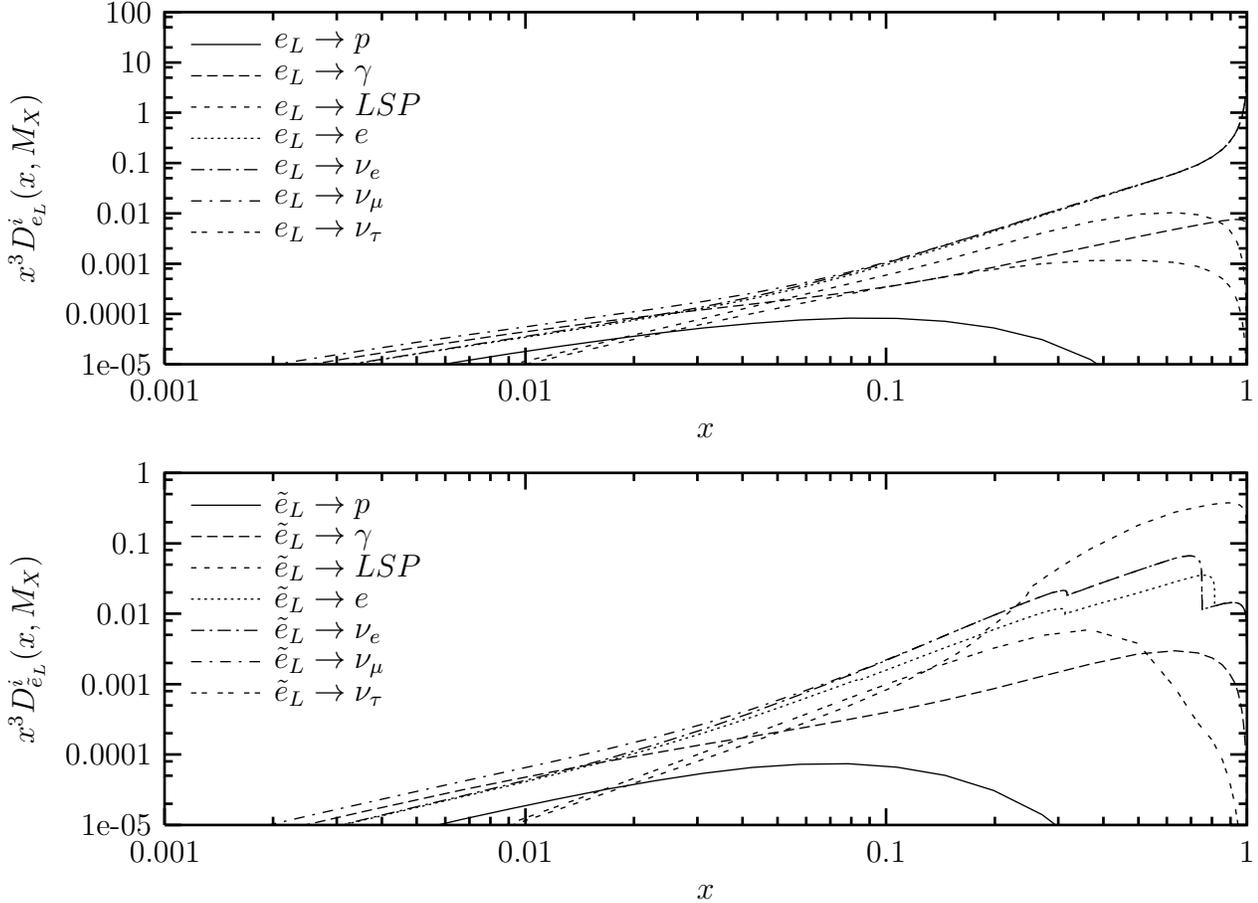

\setlength{\unitlength}{1cm}
\begin{minipage}[h]{6.5cm}
\input{Low_G_eL.tex}
\end{minipage}

\noindent
\begin{minipage}[h]{6.5cm}
\input{Low_G_eL_.tex}
\end{minipage} \hfill
\caption{Fragmentation functions of a first or second generation
$SU(2)$ doublet lepton (top) or slepton (bottom) into stable
particles. The structures in some of the curves in the lower frame
originate from 2--body decay kinematics.}
\label{leptons}
\end{figure}

Neither ultra--energetic LSPs nor neutrinos have as yet been
observed. Among the particles with large cross sections on air, at
large $x$ we observe a significantly smaller FF into protons for an
initial squark than for an initial quark. The reason is that a quark
can directly fragment into a proton, but a squark will decay first,
thereby distributing its energy over a larger number of softer
particles. Note also that the electromagnetic component (sum of FFs
into photons and electrons\footnote{In scenarios where an
electromagnetic cascade can take place during the propagation in the
extragalactic medium, the spectrum of primary electrons contributes to
the spectrum of UHE photons (see \cite{reviewSigl}).}) is always
bigger than the FF into protons. This is partly due to direct photon
emission during the early stages of the shower; this effect, which was
not included in earlier analyses, leads to a very hard component in
the photon flux. This component is about two times larger for an
initial quark than for a squark, due to the different splitting
functions. Since experiments like Haverah Park \cite{HaverahPark}
presently favor the hypothesis of hadronic primaries
\cite{UHECRconstraints}, this result could lead to problems for
top--down models. However, interactions with the intergalactic medium
will affect the electromagnetic component even more than the proton
flux. Hence top--down models could still be compatible with a primary
CR spectrum composed mostly of protons if the flux is dominated by $X$
decays at cosmological distances.

Analogous results for $SU(2$) doublet (s)leptons of the first or
second families are shown in fig.~\ref{leptons}. In case of an initial
lepton, the resulting flux for $x \gsim 0.1$ remains dominated by
leptons; due to our assumption of exact $SU(2)$ symmetry at $Q >
M_{\rm SUSY}$, electrons, $\nu_e$ and $\nu_\mu$ contribute equally
here. The flux of both LSPs and photons is much higher than that of
protons in this case. This remains true for an initial slepton, where
the LSP component is the dominant one for $x \gsim 0.2$, followed by
(charged or neutral) leptons and photons. In both cases the peak of
the proton flux (after multiplication with $x^3$) is shifted down in
energy by about a factor of 2 compared to the results of
fig.\ref{quarks}, and is reduced in size by about one order of
magnitude for large $x$. The dominance of the electromagnetic
component is therefore much more prominent for primary $X$ decays into
(s)leptons than for decays into (s)quarks.

\begin{figure}[h]
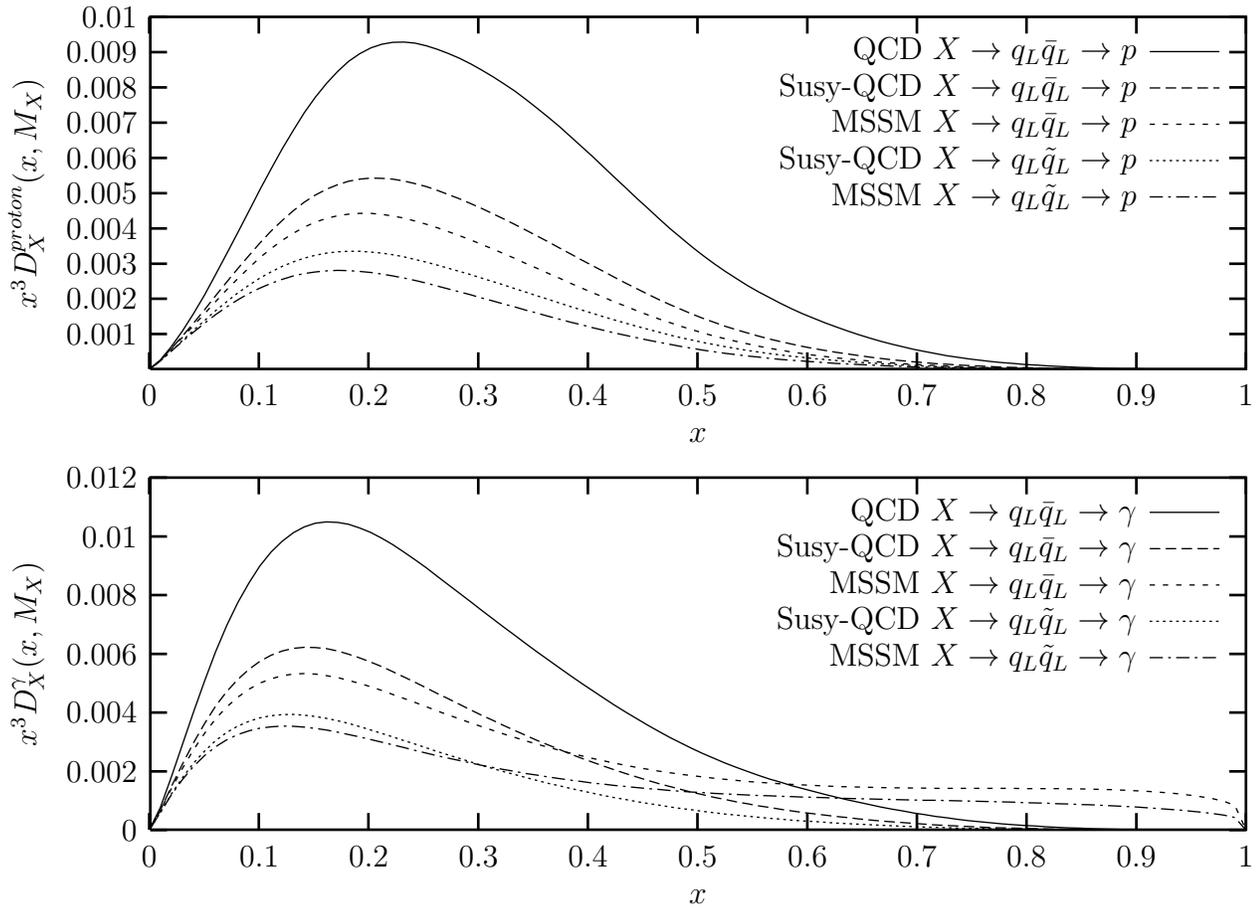

\setlength{\unitlength}{1cm}
\begin{minipage}[h]{6.cm}
\input{fragment_X_p.tex}
\end{minipage}

\noindent
\begin{minipage}[h]{6.cm}
\input{fragment_X_gamma.tex}
\end{minipage} \hfill
\caption{Fragmentation functions of $X$ in QCD, SUSY QCD
and the full MSSM for $X \rightarrow q_L\bar{q}_L$ and $X \rightarrow
\bar q_L \tilde{q}_L$; the upper (lower) frame is for fragmentation
into protons (photons).}
\label{compare}
\end{figure}

In fig.~\ref{compare} we compare the spectra of
protons and photons as predicted by QCD, SUSY QCD and the full MSSM,
for hadronic 2--body $X$ decays; in the latter two cases we show results for
$X \rightarrow q_L \bar q_L$ as well as $X \rightarrow \tilde q_L \bar
q_L$. We see that introducing superparticles into the parton shower
reduces the FF into protons by about a factor of 2 at large $x$.
Including superparticles not only opens up new channels, i.e.
introduces new splitting functions in eq.(\ref{dglap}), it also
implies a larger $SU(3)$ gauge coupling at high $Q$, which speeds up
the shower evolution through standard splitting functions. Including
in addition the full set of gauge and third generation Yukawa
interactions has relatively little impact on the proton spectrum.
Note, however, the tail of very energetic photons produced by the
electroweak gauge interactions. As already seen in fig.~\ref{quarks},
at large $x$ the flux of both protons and photons is somewhat smaller
for $X \rightarrow q$ decays than for $X \rightarrow \tilde q$.
Unfortunately it is difficult to directly compare fig.~\ref{compare}
with earlier analyses. As noted in the previous Section, we employ
somewhat different input FFs at $Q_0 = 1$ GeV; moreover, most earlier
papers did not include the contribution from sparticle decays.

We also compared the spectra of LSPs and leptons for the same set of
assumptions as in fig.~\ref{compare}. The LSP flux is very insensitive
to the inclusion of electroweak gauge and Yukawa interactions, but
changes by a factor of $\sim 5$ at $x = 0.5$ between $X \rightarrow q
\bar q$ and $X \rightarrow \bar q \tilde q$ decays. The FFs into
leptons gain a very hard component (which, however, falls
approximately linearly as $x \rightarrow 1$) once electroweak gauge
interactions are included; moreover, if $X$ decays produce a primary
squark, subsequent sparticle decays give rise to a large flux of hard
leptons at $x \sim 0.25$.

So far we have focused on the large$-x$ region of the FFs. All FFs
increase rapidly as $x$ decreases; this is not apparent in the
figures, since the increase is over--compensated by the $x^3$ scaling
factor used there. This increase towards small $x$ can have quite
dramatic effects. For example, we found that the total multiplicity of
all stable particles with $x \geq x_{min}$ can approximately be
described by $N_{tot}(x_{min}) \sim 2 \cdot x_{min}^{-0.68}$, if $X$
decays into quarks and/or squarks. This means that a single $X$ decay
will give rise to $\sim 10^5$ stable particles with energy exceeding
$5 \cdot 10^{-8} M_X$! The result for pure QCD is only about a factor
of 2 lower. The reason is that the shower evolution between $M_{\rm
SUSY}$ and 1 GeV is not affected by the presence of superparticles.
In this region QCD interactions are strongest; in the pure QCD case
this part of the shower evolution therefore contributes more to the
final multiplicity than the evolution at $Q > 1$ TeV. The large
multiplicity also implies that a full Monte Carlo simulation of this
shower \cite{Birkel} would require a very large numerical effort; this
illustrates the advantage of using fragmentation functions.

We saw earlier that at large $x$, LSPs play an important role even if
the primary $X$ decay does not involve superparticles. We found that
the integrated LSP multiplicity can be parameterized as $N_{\rm
  LSP}(x_{min}) \sim 0.65 x_{min}^{-0.32}$ for $x_{min} \lsim 0.01$,
if $X$ decays into two (s)quarks. This is somewhat smaller than the
result of ref.\cite{BerezinskyLSP}, $N_{\rm LSP}(x_{min}) \sim 0.6
x_{min}^{-0.4}$. It is also much smaller than the multiplicity of
neutrinos, which is about half of the total multiplicity discussed
above. The best hope for detecting ultra--energetic LSPs, which would
be a ``smoking gun'' signature for any top--down scenario
\cite{BerezinskyLSP}, might therefore lie in the LSPs with energies
not far below $M_X$, where the neutrino background is smallest. This
conclusion is strengthened if most UHECR originate at cosmological
distances, since then the interaction of the original protons with the
intergalactic medium will produce \cite{reviewSigl} pions, and hence
additional neutrinos; of course, these will also have energies well
below $M_X$. LSPs essentially do not interact with the intergalactic
medium. Nevertheless medium effects will be crucial for turning our
FFs into LSPs into predictions for LSP fluxes on Earth, since the flux
of protons (possibly plus photons), which is affected by the
intergalactic medium, always has to be normalized to the measured
value. Moreover, bottom--up models originally produce almost no tau
neutrinos. However, atmospheric neutrino data indicate large mixing
between $\nu_\mu$ and $\nu_\tau$, in which case the original $\nu_\mu$
flux will be distributed equally between $\nu_\mu$s and $\nu_\tau$s
after at most a few kpc of propagation. Therefore a very energetic
$\nu_\tau$ flux can unfortunately by itself not discriminate between
these kinds of models.

We also saw in figs.~\ref{quarks} and \ref{leptons} that a flux of
neutrinos with energies above those of the most energetic protons is a
generic feature of top--down models. This signal is especially
pronounced if primary $X$ decays produce squarks and/or (s)leptons; it
will be strengthened by interactions of the protons with the medium,
which move the protons to lower energies. In contrast, in bottom--up
models one expects the neutrino flux to cut off at significantly lower
energies than the original proton flux. However, even in this kind of
model the neutrino flux might extend to higher energies than the
observed proton flux once the protons have traveled for a few tens of
Mpc. A very hard component in the neutrino flux can therefore only be
considered to be a signature for top--down models if it can be shown
that most UHECR are produced ``locally'', or if the neutrino spectrum
is as shown in the upper fig.~\ref{leptons}, with a sharp peak at the
highest energy.

%%%%%%%%%%%%%%%%%%%%%%%%%%%%%%%%%%%%%%%%%%%%%%%%%
\section{Summary and Conclusions}
\label{sec:conclusion}
%%%%%%%%%%%%%%%%%%%%%%%%%%%%%%%%%%%%%%%%%%%%%%%%%

In this paper we presented a new and quite complete algorithm for
treating the decay of a superheavy $X$ particle in the framework of
the MSSM. We improved previous analyses by including the full set of
MSSM gauge and third generation Yukawa interactions, and by carefully
modeling the decays of heavy sparticles and particles.  We applied
this algorithm to the ``top--down'' solution of the problem of
UHECR. We gave the results in terms of fragmentation functions for any
possible decay product of $X$ into stable particles, and studied the
consequences for a few primary decay modes of the $X$ particle itself.
An important result of this complete study, compared to a simplified
SUSY QCD treatment, was the prediction of sizable fluxes of leptons
and photons at energies above the peak of the proton spectrum. We also
discussed possible ways to distinguish between bottom--up and
top--down scenarios, and concluded that the cleanest signal would
probably come from the detection of LSPs with energies above the peak
of the proton spectrum. Of course, detailed analyses are required
before we can conclude that these LSPs are indeed detectable on the
background of neutrinos.

So far the only UHECR that have been detected are hadrons, or
perhaps photons. If propagation effects can be neglected,
our results can be used directly to fit the mass $M_X$ of the
primary. If $X$ decays primarily into (s)quarks, the result of such a
fit would presumably resemble that of \cite{Sarkar:2001}, if we assume
that all events with energy above a few times $10^{19}$ eV originate
from $X$ decay. This would favor a relatively ``low'' $M_X \sim
10^{12}$ GeV, well below the GUT scale. On the other hand,
ref.\cite{Fodor} found a much higher value of $M_X$, near the GUT
scale, if $X$ decays at cosmological distances. However,
ref.\cite{Fodor} used a relatively crude model for $X$
decay.\footnote{For example, their input FFs violate energy
conservation badly.} It might be worthwhile to re--do this analysis,
which requires the careful treatment of propagation effects. Finally,
there is some evidence that there are two different sources for
post--GZK events. For example, it has been claimed \cite{Clusters} that
the number of ``doublet'' and ``triplet'' events (which originate from
a small patch in the sky) is too high to be compatible with an
essentially 
isotropic distribution of sources, which is the most plausible
assumption for a top--down model\footnote{However, clustering of
events can be explained in top--down models with ``clumpy halo''
\cite{Blasi:2000,Blasi:2001}}. Current statistics is poor, but
there is some indication that this clustering of events occurs only at
energy below $10^{20}$ eV. Moreover, most of the experiments (Agasa,
Yakutsk, and the Fly's eye collaboration) indicate that there might be
a cut--off in the spectrum, at an energy $\sim 4 \cdot 10^{19}$ eV (see
\cite{Yoshida} for a review). This might also hint towards a
two--component explanation for the events before and above the
cut--off \cite{Bird:1993}. If so, it might be best to only use events
with energy well above the GZK cutoff, say with $E > 10^{20}$ eV, in
the fit of $M_X$; in that case the small number of events would
presumably allow large values of $M_X$ even if most $X$ decays are
``local''. In any case, when expressed in terms of the dimensionless
scaling variable $x = 2 E/M_X$, the fragmentation functions only
depend logarithmically on $M_X$.

If $M_X \gg 10^{21}$ eV the events that have been observed so far
would all have $x \ll 1$. The fluxes at small $x$ are much higher than
in the large $x$ region. Even if the observed UHE events come from the
large $x$ region, i.e. if $M_X \sim 10^{21}$ eV, care has to be taken
not to over--produce particles at lower energies. In particular,
stringent limits exist on the fluxes of photons in the TeV energy
region, and on neutrinos in the multi--TeV region. The interpretation
of these limits in the framework of a given top-down model again
depends on where $X$ decays occur, i.e. if propagation effects are
important or not. We note here that in our case the evolution equation
(\ref{dglap}) can still be applied at $x$ as small as
$10^{-7}$. Coherence effects, which give rise to the ``MLLA''
description of the parton shower, become large \cite{Marchesini} at
yet smaller values of $x \sim \sqrt{Q_h/M_X}$, where $Q_h$
characterizes the hadron mass scale where strong interactions become
non--perturbative. We also checked that the precise form of our
small$-x$ extrapolation of the nonperturbative FF's is not important,
since the small$-x$ behavior of the final FF's is essentially
determined by the perturbative evolution. However, perturbative
higher order corrections might be sizable at small $x$; this effect is
currently being investigated.

To summarize, we presented a first treatment of superheavy particle
decay that includes all relevant particle physics effects at the
leading--log order of perturbation theory; in particular, for the
first time we were able to account for the entire energy released in
the decay. This provides a tool which can be used for detailed tests
of the hypothesis that the most energetic cosmic rays originate from
the decay of ultra--massive particles. If true, this would give us
experimental access to energies well beyond the range that can ever be
tested by experiments working at Earth--based colliders.

%%%%%%%%%%%%%%%%%%%%%%%%%%%%%%%%%%%%%%%%%%%%%%%%%
\section*{Acknowledgements}
%%%%%%%%%%%%%%%%%%%%%%%%%%%%%%%%%%%%%%%%%%%%%%%%%

We would like to thank S. Sarkar for a very interesting discussion.
This work  was supported in part by the SFB 375 Astro--Teilchenphysik
of the Deutsche Forschungsgemeinschaft.

%--------------------------- References -------------------------------

\bibliographystyle{utphys}
\bibliography{references}

\clearpage
\noindent

\end{document}